\documentclass[letterpaper,10pt]{article}
\usepackage{osameet2}

\usepackage{wrapfig}
\usepackage[labelsep=period,justification=centering,font=footnotesize]{caption}
\usepackage{graphicx}
\usepackage{float}
\usepackage{pgfplots}
\usepackage{tikz}
\usetikzlibrary{plotmarks}
\usepackage{pgfplots}
\usepgfplotslibrary{colorbrewer}
\pgfplotsset{compat=1.12}
\pgfkeys{/pgf/number format/.cd,1000 sep={\,}}

\usepackage{amsmath,amssymb,mathtools}

\usepackage[pdftex,colorlinks=true,bookmarks=false,citecolor=blue,urlcolor=blue]{hyperref} 

\newcommand\authormark[1]{\textsuperscript{#1}}

\begin{document}

\title{120~Tbit/s Transmission over Single Mode Fibre using Continuous 91~nm Hybrid Raman-EDFA Amplification}

\author{L~Galdino\authormark{1,*},
A~Edwards\authormark{2},
M~Ionescu\authormark{2},
J~James\authormark{2},
W~Pelouch\authormark{3},
E~Sillekens\authormark{1},
D~Semrau\authormark{1},
D~Lavery\authormark{1},
R~I~Killey\authormark{1}, 
S~Barnes\authormark{2},
P~Bayvel\authormark{1},
and S~Desbruslais\authormark{2}}

\address{\authormark{1}Optical Networks Group, Dept. of Electronic \& Electrical Engineering, University College London, UK \\
\authormark{2}Xtera, Bates House, Church Road, Harold Wood, Essex, UK\\
\authormark{3}{Xtera, 500 West Bethany Drive,
Allen,Texas, USA}}

\email{\authormark{*}l.galdino@ucl.ac.uk}

\begin{abstract}
Hybrid distributed Raman-EDFA amplifiers, with a continuous 91~nm gain bandwidth, are used to enable a record single mode fibre transmission capacity of 120.0~Tbit/s using 312$\times{}$35~GBd DP-256QAM over 9$\times{}$70~km spans.
\end{abstract}


\section{Introduction}

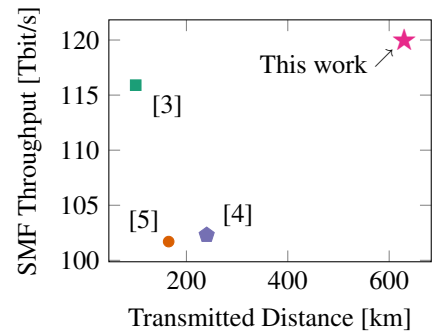
\begin{wrapfigure}[12]{r}{0.35\textwidth}
\vspace{-0.45cm}

\begin{tikzpicture}
\begin{axis}[width=1\linewidth,
height=4.8cm,
xlabel={Transmitted Distance [km]},
ylabel={SMF Throughput [Tbit/s]}
]
\addplot[mark=square*,Dark2-A] coordinates {(100,115.9)};
\node[anchor=north west] at (axis cs:100,115.9) {\cite{100nm}};

\addplot[mark=*,Dark2-B] coordinates {(165,101.7)};
\node[anchor=south east] at (axis cs:165,101.7) {\cite{Sano2}};

\addplot[mark=pentagon*,mark size=3pt,Dark2-C] coordinates {(240,102.3)};
\node[anchor=south west] at (axis cs:240,102.3) {\cite{Sano}};

\addplot[mark=text,text mark={$\bigstar$},Dark2-D] coordinates {(630,120)};
\node[anchor=north east] at (axis cs:630,120) {This work $^\nearrow$};
\end{axis}
\end{tikzpicture}
   \vspace{-0.4cm}
\caption{\small Capacity records in a single-mode fibre}
\label{fig:capacityrecord}
\end{wrapfigure}

Fundamental limits of an optical communication system are imposed by a combination of noise from the transceiver subsystem, optical amplifier noise and optical fibre nonlinearity. Advances in digital signal processing (DSP), modulation formats \cite{SE}, high speed electronics, transmission fibres, and broadband optical amplification, have resulted in recent experimental single-mode fibre (SMF) capacity increases. Recent years have seen several landmark transmission results using single mode fibre, with demonstrations achieving throughputs exceeding  70~Tbit/s over transatlantic distances \cite{70Tbps} and an overall throughput record of 115~Tbit/s over 100~km \cite{100nm}.

With the exception of \cite{100nm} \cite{Sano}, previous record demonstrations have relied on C+L-band EDFA transmission, in which a spectral gap between the two bands was required in order to use EDFAs in each transmission window. In \cite{100nm}, a continuous-band 100~nm amplifier was developed based on a semiconductor optical amplifier (SOA), enabling an SMF capacity record of 115.9~Tbit/s over 100~km. 
Although the bandwidth is impressive, SOAs have a relatively high noise figure compared with distributed Raman amplifiers, and so system performance decreases rapidly with distance. 

In this work, we demonstrate record throughput of 120.0~Tbit/s transmitted over single mode fibre using a 91~nm continuous gain bandwidth hybrid distributed Raman-EDFA (HRE), designed as a prototype for wideband amplification. As highlighted in Fig.~\ref{fig:capacityrecord}, the data throughput and distance are both increased over the previous capacity record, with an extra 4~Tbit/s capacity over more than 6 times longer transmission distance.

\begin{figure}[tb]
  \centering  \includegraphics[width=0.85\textwidth]{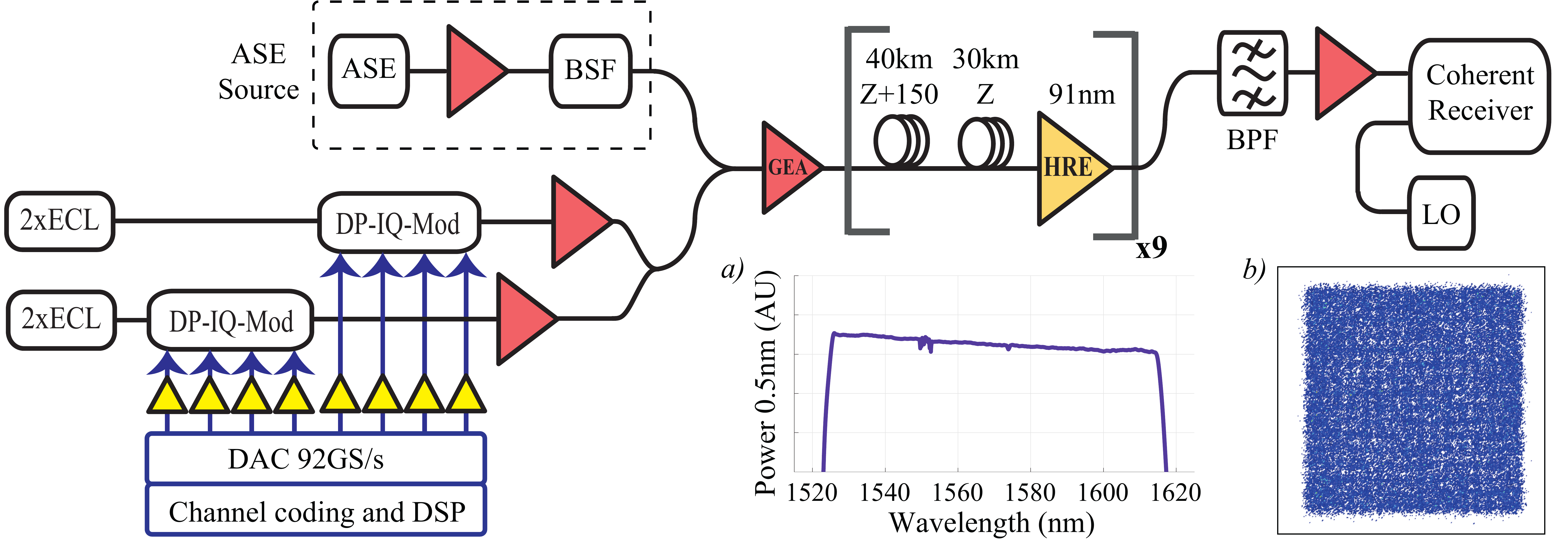}
\caption{Schematic of the transmission experiment. Inset a) signal spectrum after the GEA (the modulated channels are centered at 1551~nm. Inset b) 256QAM constellation measured in back-to-back configuration.  }
\label{fig:setup}

\end{figure}

\section{Experimental Configuration}

The experimental configuration used for this work is shown in Fig.~\ref{fig:setup}.  Four carriers, spaced at 35.5~GHz, were connected to two independent dual-polarization IQ optical modulators, each driven by four 92~GS/s digital-to-analogue converters (DACs) to generate four odd/even channels. A digital root-raised cosine (RRC) filter with 0.01 roll-off was used to spectrally shape the signals and pre-emphasis was applied to overcome the electrical response of the transmitter components. The channels were generated at carrier frequencies which were tuned across the range 185.5405 to 196.5810~THz, 
allowing the measurement of 312$\times{}$35~GBd dual-polarization 256-\textit{ary} quadrature amplitude modulation (DP-256QAM) channels, covering the spectrum from 1524.9 to 1615.9~nm.

The modulated channels were amplified using a pair of 97~nm bandwidth discrete Raman amplifiers with 12.5~dB gain and combined with wideband amplified spontaneous emission (ASE) noise, which emulated co-propagating channels over the entire  
transmitted bandwidth. The ASE noise, with continuity across the entire 97~nm bandwidth, was generated by a pair of discrete Raman amplifiers to achieve a total output power of 22~dBm within the required 91nm bandwidth, following which, a band stop filter (BSF) was used to create a notch in the ASE within which the modulated channels were positioned. 
The validity of using ASE noise to emulate agressor channels was verified in \cite{ASE}, showing that this technique provides a conservative measure of system performance. 
This technique has been used in \cite{100nm} and \cite{trans-atlantic} to estimate transmission system capacity. 
A gain equalizing amplifier (GEA) with a continuous 91.04~nm bandwidth was used to amplify and spectrally shape (SS) the combined ASE and modulated channels. The combined SS-ASE and modulated channels occupied a total useable bandwidth of 11.0758~THz (91.04~nm) with a total output power of 22~dBm. The optical spectrum after the GEA is shown as inset (a) in Fig.~\ref{fig:setup}. A power tilt of -2~dB across the bandwidth was applied to optimize channel performance, taking into account the wavelength-dependent noise figure of the transmission line amplifiers. The back-to-back 256QAM constellation is illustrated in the inset (b) of Fig.~\ref{fig:setup}.

The transmission link comprised a straight-line of 9~spans, with 70~km of single-mode fibre and a hybrid distributed hybrid Raman-EDFA amplifier (HRE) in each span. Each HRE provided continuous gain from 1525~nm to 1616~nm and used two counter-propagating pumps at 1427 and 1495~nm with output powers of 300~mW and 310~mW into the transmission fibre, delivering a total power of 19.5~dBm to the EDFA stage. The single EDFA stage boosted the signal to a total output power of 22~dBm, and included a 91~nm gain flattening filter (GFF) to equalize the gain across the entire HRE bandwidth. Each 70 km fibre span comprised two fibre types; the first part of the hybrid span used 40 km of Sumitomo Z+150 fibre with an average attenuation of {0.148~dB/km} and an effective core area of {$149$~{\textmu}m$^{2}$}. The second part of the span used 30~km of Sumitomo Z~fibre with an average attenuation of {0.16~dB/km} and an effective core area {$81$~{\textmu}m$^{2}$}. The second segment of the hybrid span used a lower effective area to increase the Raman gain in the second part of the span for a given pump power. 

At the receiver, a bandpass filter (BPF) with a 40~GHz bandwidth was used to filter the channel under test. The BSF and the transmitter BPF were moved together in order to create a notch for the channel under test (channel 2 in the 4-channel group) and exclude the out-of-band noise from the receiver.  The coherent detection was carried out using a phase- and polarization-diverse coherent receiver incorporating 70~GHz bandwidth photodetectors, and the signal was digitized using a real-time oscilloscope with 63~GHz bandwidth, sampling at 160 GSa/s. Digital signal processing was performed as described in \cite{ASE}, which included matched filtering, single step chromatic dispersion compensation, a 21-tap blind adaptive equalizer, frequency offset compensation and decision directed carrier phase estimation. To quantify the system performance all 312 measured
channels were individually decoded with soft decision forward error correction (SDFEC) and were accepted if the post-SDFEC was below the assumed outer hard decision-FEC (HDFEC) threshold. For the FEC,
12~rate adapted low density parity check (LDCP) codes from the DVB-S2X standard were implemented.
The assumed outer HD-FEC with a bit error rate (BER) threshold of $3\times 10^{-4}$ allows for a BCH code with $0.5\%$ overhead. System performance was also quantified in terms of information rates.  Signal-to-noise ratio (SNR) was estimated for every channel and the generalised mutual information (GMI) was calculated on a per channel basis.

\section{Results}

Fig.\ref{fig:SE} plots the information rate of all 312~x~35~GBd DP-256QAM channels, including the 0.5~GHz channel spacing. The green

diamonds show the Shannon limit for the received SNR given by $\log_2(1+\mathrm{SNR})$ summed over both polarisations. The total received SNR was evaluated as the ratio between the variance of the transmitted symbols $E[|X|^2]$ and the variance of the noise $\sigma^2$, where $\sigma^2=E[|X-Y|^2]$ and $Y$ represents the received symbols. The blue triangles depict the pre-FEC GMI calculated using received log-likelihood ratios. The post-FEC rate after inner and outer FEC is plotted by the red circles. A mean penalty of 0.94~bit/symbol between the GMI (11.62~bit/symbol) and the Shannon limit (12.56~bit/symbol) is due to the use of non-optimal finite constellation and bit labeling. For the net rate after FEC, the occupied spectrum yields a net bit rate of 10.99~bits/symbol providing a record single mode fibre capacity of 120.0~Tbit/s.

Fig.~\ref{fig:BER} shows the pre- and post-SDFEC BER for all 312 channels. The channels were decoded using 12~rate adapted LDPC codes implemented from the DVB-S2x. An outer BCH HD-FEC code with a bit error rate (BER) threshold of $3\times 10^{-4}$  with $0.5\%$ overhead was assumed. All 312 channels were measured, confirming the total net throughput of 120.0 Tbit/s. Fig.~\ref{fig:LDPC} shows the rates used for the SDFEC after which all channels were below the BER threshold for the outer HDFEC. The bulk of the codewords have a low spread, 293 out of 312 are between 40-50\% FEC overhead. Reducing the number of 7 code rates applied was found to reduce the total net data throughput to 119.0~Tbit/s.

%

\begin{figure}[tb]
\begin{minipage}[t][][t]{0.4\linewidth}
\centering
\begin{tikzpicture}
\begin{axis}[
width = 1.05\textwidth,
height = 6cm,
xlabel={Channels [nm]},
ylabel={Throughput [bit/symbol]},
grid=major,
legend reversed=true,
legend style={
	font=\footnotesize,
    at={(0.5,0.05)},
    anchor=south,
    },
legend cell align=left,
] 

\addplot[Set1-A,only marks,mark=o,mark size=2pt] table[col sep=comma,x expr=299792458/\thisrow{Fc}/1000,y expr=16/1.005*\thisrow{coderate}]{postfec.csv};
\addlegendentry{Net rate after FEC}

\addplot[Set1-B,only marks,mark=triangle,mark size=2pt] table[col sep=comma,x expr=299792458/\thisrow{Fc}/1000,y expr=\thisrow{GMI}]{postfec.csv};
\addlegendentry{GMI}

\addplot[Set1-C,only marks,mark=diamond,mark size=2pt] table[col sep=comma,x expr=299792458/\thisrow{Fc}/1000,y expr=2*log2(1+10^(\thisrow{SNR}/10)]{postfec.csv};
\addlegendentry{$\log_2(1+\mathrm{SNR}) $}

\end{axis}
\end{tikzpicture}
\caption{Throughput per channel over 2 polarisations after 630 km.}

\label{fig:SE}

\end{minipage}%
\begin{minipage}[t][][t]{0.4\linewidth}
\centering
\begin{tikzpicture}
\begin{semilogyaxis}[
width = 1\textwidth,
height = 6cm,
xlabel={Carrier wavelength [nm]},
ylabel={BER},
grid=major,
legend reversed=true,
legend style={
	font=\footnotesize,
    at={(0.5,0.75)},
    anchor=base,
    },
ymin = 1e-8,
xmin=1515,xmax=1625,
legend cell align=left,
legend columns=2,
log origin=infty,
] 

\addplot[
	Set1-A,
   	only marks,
	mark=text,
    text mark={\tiny $\square$},
    ] table[col sep=comma,x expr=299792458/\thisrow{Fc}/1000,y expr=\thisrow{postFECBER}]{postfec.csv};
\addlegendentry{Post-SDFEC};

\addplot[
	Set1-A,
    only marks,
    mark=text,
    text mark={\tiny $\mathrlap{_\downarrow}\square$},
    forget plot,
    y filter/.code={\pgfmathparse{ifthenelse(rawy==0,-16,nan)}},
] table[col sep=comma,x expr=299792458/\thisrow{Fc}/1000,y=postFECBER]{postfec.csv};

\addplot[Set1-B,only marks,mark=+] table[col sep=comma,x expr=299792458/\thisrow{Fc}/1000,y expr=\thisrow{preFECBER}]{postfec.csv};
\addlegendentry{Pre-FEC};

\addplot[black,dashed,thick,forget plot,domain=1515:1630] {3e-4};
\node[anchor=south west,font=\footnotesize] at (axis cs:1520,3e-4) {$0.5\%$ HD-FEC threshold};

\draw[<-,thick] (axis cs:1570,0.7e-7) -- (axis cs:1580,0.25e-7) node[anchor=west,align=left,style={font=\footnotesize}] {no errors \\[-0.3em] detected};

\end{semilogyaxis}
\end{tikzpicture}
\caption{Pre-FEC and Post-SDFEC bit error rate for all 312 channels. Arrows indicate channels where no errors were observed after post-SDFEC.}
\label{fig:BER}
\end{minipage}%
\begin{minipage}[t][][t]{0.18\linewidth}
\centering
\begin{tikzpicture}
\begin{axis}[
width=1.2\textwidth,
height=6cm,
xlabel={LDPC OH [\%]},
ylabel={Number of Channels},
ymin=0,
xmode=log,
log ticks with fixed point,
xtick={40,60,100}
]
\addplot[ybar,Set1-B,fill,bar width=0.2pt] table[col sep=comma,x expr=(1/\thisrow{x}-1)*100]{
x,y
0.45,1
0.556,3
0.642,15
0.667,10
0.675,12
0.68,24
0.686,36
0.691,36
0.696,59
0.702,48
0.708,52
0.714,16
};
\end{axis}
\end{tikzpicture}
\caption{LDPC overhead usages.}
\label{fig:LDPC}
\end{minipage}

\vspace{-0.8cm}
\end{figure}

\section{Conclusion}
A record capacity of 120~Tbit/s in a single mode optical fibre was realised through the development of a wide-bandwidth hybrid EDFA-Raman amplification scheme. In contrast to the most recent record capacity demonstration \cite{100nm}, this technique offers distributed gain, and so inherently maintains a high SNR over long transmission distances and the relatively long 70~km spans used in this work. In contrast to EDFA-based C+L band transmission systems, the continuous 91~nm gain bandwidth offered by hybrid amplification means that no spectrum is wasted. 

Finally, we note that this demonstration has been made using mature signal technologies (standard square QAM modulation and binary LDPC codes) and is, thus, eminently and readily realisable for commercial deployment.

\vspace{0.4cm}
\noindent
\footnotesize{\textit{Support from UK EPSRC UNLOC programme and the Royal Academy of Engineering under the Research Fellowships programme is gratefully acknowleged. The authors thank Oclaro for the modulators and Nick Hepden for the heroic effort in constructing the hybrid amplifiers used in this work.}}

\end{document}